\documentclass{article}
\usepackage{ijcai25}

\usepackage{times}
\usepackage{soul}
\usepackage{url}
\usepackage[hidelinks]{hyperref}
\usepackage[utf8]{inputenc}
\usepackage[small]{caption}
\usepackage{graphicx}
\usepackage{amsmath}
\usepackage{amsthm}
\usepackage{booktabs}
\usepackage{algorithm}
\usepackage{algorithmic}
\usepackage[switch]{lineno}
\usepackage{mydef}

\title{Graph Neural Networks for Databases: A Survey}

\author{
Ziming Li$^1$
\and
Youhuan Li$^{1,*}$\and
Yuyu Luo$^2$\and
Guoliang Li$^3$\and
Chuxu Zhang$^4$
\affiliations
$^1$Hunan Univerisity, 
$^2$The Hong Kong University of Science and Technology (Guangzhou)\\
$^3$Tsinghua University, 
$^4$University of Connecticut\\
\emails
\{zimingli, liyouhuan\}@hnu.edu.cn,
yuyuluo@hkust-gz.edu.cn, \\
liguoliang@tsinghua.edu.cn,
chuxu.zhang@uconn.edu
}

\begin{document}

\maketitle

\begin{abstract}
Graph neural networks (GNNs) are powerful deep learning models for graph-structured data, demonstrating remarkable success across diverse domains. Recently, the database (DB) community has increasingly recognized the potentiality of GNNs, prompting a surge of researches focusing on improving database systems through GNN-based approaches. However, despite notable advances, There is a lack of a comprehensive review and understanding of how GNNs could improve DB systems.  
Therefore, this survey aims to bridge this gap by providing a structured and in-depth overview of GNNs for DB systems. Specifically, we propose a new taxonomy that classifies existing methods into two key categories: (1) Relational Databases, which includes tasks like performance prediction, query optimization, and text-to-SQL, and (2) Graph Databases, addressing challenges like efficient graph query processing and graph similarity computation. We systematically review key methods in each category, highlighting their contributions and practical implications.  
Finally, we suggest promising avenues for integrating GNNs into Database systems.

\end{abstract}
\section{Introduction}
\begin{figure}[t]
	\centering 
	\resizebox{\linewidth}{!}
	{
		\includegraphics[width=\linewidth]{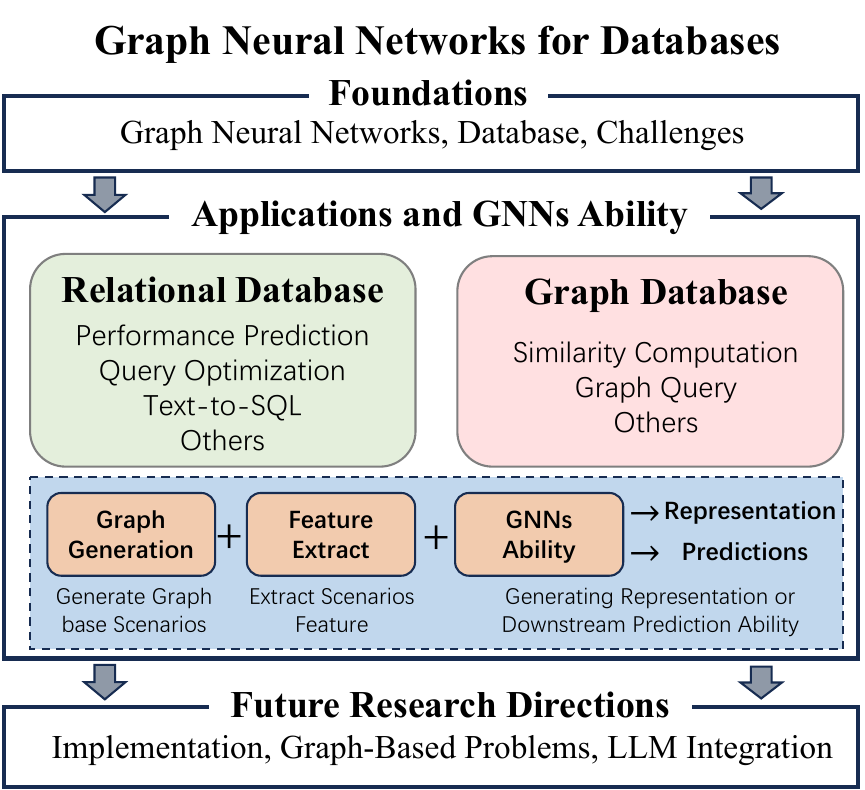}
	}
	\vspace{-0.25in}
	\caption{Overview of Survey.}
	\vspace{-0.1in}
	\label{fig:narrative}
\end{figure}

\tikzstyle{leaf}=[draw=hiddendraw,
    rounded corners,minimum height=1em,
    fill=mygreen!40,text opacity=1, align=center,
    fill opacity=.5,  text=black,align=left,font=\scriptsize,
    inner xsep=3pt,
    inner ysep=1pt,
    ]
\tikzstyle{middle}=[draw=hiddendraw,
    rounded corners,minimum height=1em,
    fill=output-white!40,text opacity=1, align=center,
    fill opacity=.5,  text=black,align=center,font=\scriptsize,
    inner xsep=3pt,
    inner ysep=1pt,
    ]
\newcommand{\twthree}{8.9em}
\newcommand{\twfour}{28.2em}
\begin{figure*}[ht]
\centering
\begin{forest}
  for tree={
    forked edges,
    grow=east,
    reversed=true,
    anchor=base west,
    parent anchor=east,
    child anchor=west,
    base=middle,
    font=\scriptsize,
    rectangle,
    line width=0.7pt,
    draw=output-black,
    rounded corners,align=left,
    minimum width=2em,
    s sep=5pt,
    inner xsep=3pt,
    inner ysep=1pt,
  },
  where level=1{text width=4.5em}{},
  where level=2{text width=6em,font=\scriptsize}{},
  where level=3{font=\scriptsize}{},
  where level=4{font=\scriptsize}{},
  where level=5{font=\scriptsize}{},
  [Graph Neural Network for Databases, middle,rotate=90,anchor=north,edge=output-black
    [Relational Databases, middle, edge=output-black,text width=8.5em
        [Performance Prediction, middle, text width=\twthree, edge=output-black
            [Neo~\cite{traditionaldb-plan-Neo}{, }GPredictor~\cite{traditionaldb-performance-GPrecdictor}{, }Bao~\cite{bao} \\Stage~\cite{tradition-predict-stage}{, }Tosure~\cite{tradition-predict-tosure}{, }BRAD~\cite{tradition-prediction-brad}, leaf, text width=\twfour, edge=output-black]
        ]
        [Query Optimization, middle, text width=\twthree, edge=output-black
            [RTOS~\cite{traditionaldb-join-RTOS}{, }OpenGauss~\cite{traditional-join-gauss}{, }JOGGER~\cite{traditionaldb-join-JOGGER}\\ SOAR~\cite{traditionaldb-join-soar}{, }RGOS~\cite{traditional-query-RGOS}{, }LOGER~\cite{traditionaldb-plan-LOGER} \\
            GnnMV~\cite{traditionaldb-View-GnnMV}{, }AutoCE~\cite{traditionaldb-performance-AutoCE}{, }GCE~\cite{tradition-predict-GCE}, leaf, text width=\twfour, edge=output-black]        
        ]
        [Text-to-SQL, middle, edge=output-black, text width=\twthree
            [GNNSQL~\cite{sql-bogin}{, }GLOBAL-GNN~\cite{sql-globalgnn}{, }SADGA~\cite{sql-sadga}\\LGESQL~\cite{sql-lgesql}{, }ShadowGNN~\cite{sql-shadowgnn}{, }IST-SQL~\cite{sql-ist-sql}\\ISESL-SQL~\cite{sql-ISESL-SQL}{, }S$^2$SQL~\cite{sql-s2sql}{, }Graphix-T5~\cite{sql-t5}\\TFFSQL~\cite{sql-tffsql}{, }GRL-SQL~\cite{sql-GRL-SQL}, leaf, text width=\twfour, edge=output-black]
        ]
        [Others, middle, text width=\twthree, edge=output-black
            [Decima~\cite{tradition-predict-Decima}{, }SmartIndex~\cite{tradition-predict-smartindex}{, }Grep~\cite{traditional-partition-grep}, leaf, text width=\twfour, edge=output-black]
        ]
    ]
    [Graph Databases, middle, edge=output-black, text width=8.5em
        [Similarity Computation, middle, text width=\twthree, edge=output-black
            [SimGNN~\cite{graphdb-gsc-gsc-simgnn}{, }GMN~\cite{graphdb-GMN}{, }GraphSim~\cite{graphdb-GraphSim}\\GHashing~\cite{graphdb-gsc-gsc-ghashing}{, }Noah~\cite{graphdb-gsc-ged-noah}{, }GENN-A*~\cite{graphdb-gsc-ged-GENN-A*}\\H$^2$MN~\cite{graphdb-gsc-gsc-h2mn}{, }GraphOTSim~\cite{graphDB-gsc-gsc-graphotsim}{, }GREED~\cite{graphDB-gsc-ged-greed}\\ERIC~\cite{graphDB-gsc-gsc-areg}{, }GEDGNN~\cite{graphdb-gsc-ged-gedgnn}{, }MATA$^*$~\cite{graphDB-gsc-ged-mata*}, leaf, text width=\twfour, edge=output-black]
        ]
        [Graph Query, middle, text width=\twthree, edge=output-black
            [NeuroMatch~\cite{graphDB-gq-sm-neuromatch}{, }NSIC~\cite{graphDB-gq-sc-rgin}{, }LSS~\cite{graphDB-gq-sc-alss_sig} \\ Duong~\cite{graphDB-gq-sm-chi}{, }RL-QVO~\cite{graphDB-gq-sm-rlqvo}{, }NeurSC~\cite{graphDB-gq-sc-neursc}\\ISONET~\cite{graphDB-gq-sm-isonet}{, }Count-GNN~\cite{graphDB-gq-sc-countgnn}{, }LearnSC~\cite{graphDB-gq-sc-learnsc}\\GNN-PE~\cite{graphDB-gq-sm-gnnpe}{, }LRP~\cite{graphDB-gq-sc-mpnns}{, }DeSCo~\cite{graphDB-gq-sc-DeSCo}, leaf, text width=\twfour, edge=output-black]        
        ]
        [Others, middle, text width=\twthree, edge=output-black
            [GCNSplit~\cite{graphDB-others-partition-gcnsplit}{, }HRQE~\cite{kg-ce-hrpe}{, }GNCE~\cite{kg-ce-gnce}, leaf, text width=\twfour, edge=output-black]
        ]
    ]
  ]
\end{forest}
\caption{A Taxonomy of GNNs for Databases. The technologies of GNNs for DBs can be categorized into two types based on application scenarios: relational databases and graph databases.}
\label{fig:taxonomy}
\end{figure*}
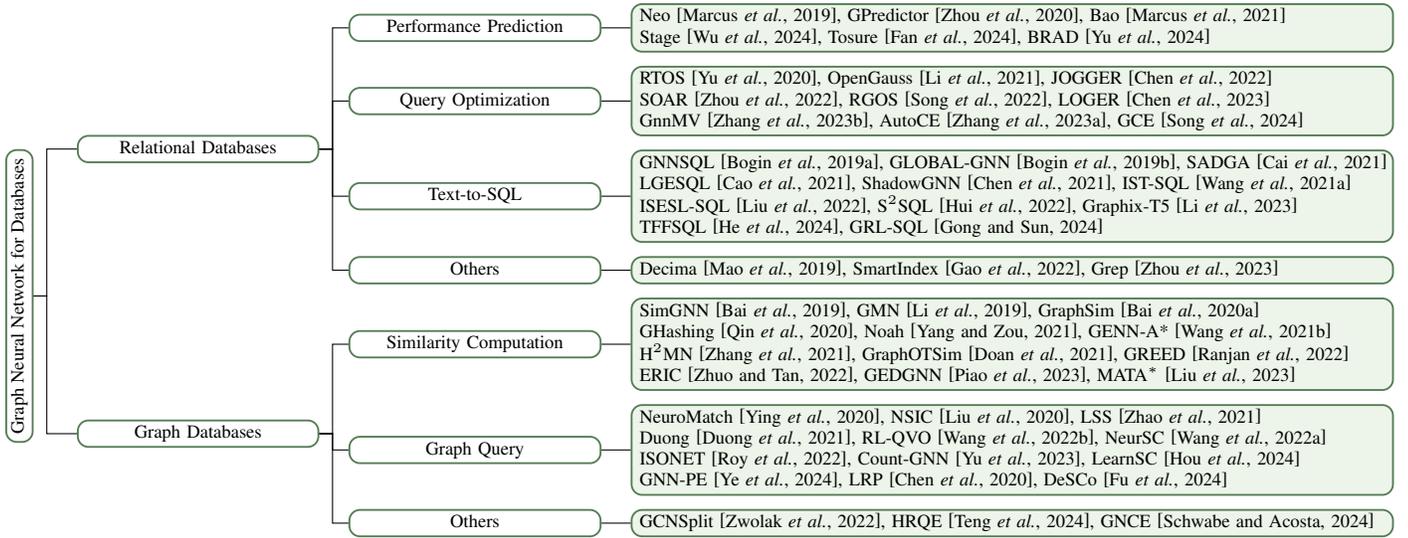
  
Graph Neural Networks (GNNs)~\cite{gcn,gat,RGCN,HGNN} have achieved remarkable success across various domains by leveraging recursive message-passing and aggregation mechanisms, which enables effective solutions for downstream tasks such as node classification and graph clustering.
In the field of databases (DB) - a well-established field with decades of research, the potential of GNNs to optimize complex DB problems has recently gained significant attention. GNNs have been applied to various challenges in DB, including query optimization, performance prediction, and graph query processing. Despite these advances, there is still a lack of comprehensive surveys that systematically categorize and summarize these contributions.

To fill this gap, this paper provides a comprehensive review of GNNs for DB. As illustrated in Fig.~\ref{fig:narrative}, we begin by introducing foundations of GNNs for DB and propose a taxonomy to categorize existing techniques. Following this, we discuss the key challenges associated with harnessing GNNs' capabilities in DB. We then delve into a detailed review of these techniques, focusing on three key questions: (1) What types of graph are utilized in specific scenarios?
(2) How are features extracted from these graphs?
(3) What roles do GNNs play in these problems - do they primarily generate data representations for downstream tasks, support downstream prediction, or perform both functions?
For each category, we include a summary section that highlights how GNNs enhance the efficiency and capability of DB.
Finally, we outline research directions to inspire future exploration of GNNs for DB.
\begin{itemize}
    \vspace{-0.05in}
    \item To the best of our knowledge, this is the first comprehensive survey that focuses exclusively on the applications of GNNs in DB systems.
    \vspace{-0.05in}
    \item We systematically review and categorize existing methods, providing insights into how graphs are constructed, how features are initialized, and the specific roles that GNNs play in each study.
    \vspace{-0.05in}
    \item We discuss promising avenues for future research at the intersection of GNNs and DB.
\end{itemize}

\section{Foundations of GNNs for Databases}
In this section, we introduce the foundations of GNNs for DB, including their definitions. Then, we discuss the challenges of integrating GNNs with DB, emphasizing aspects that necessitate careful consideration.

\subsection{Graph Neural Networks}
Graph Neural Networks (GNNs) use a message-passing paradigm over graphs to aggregate information across nodes and edges. 
A graph is defined as $ G = (\mathcal{V}, \mathcal{E}, X) $, where $ \mathcal{V} $ is the set of nodes, $ \mathcal{E} $ is the set of edges, and $ X $ indicate node features. GNNs iteratively update the embeddings of each node $ v \in \mathcal{V} $ and merge these data with the current embedding of the node. The node embedding update process on the $ (l+1) $-th layer is expressed as:
\[
h_v^{l+1} = \text{COM} \left( h_v^l, \text{AGG} \left( \{ h_u^l \mid \forall u \in \mathcal{N}_v \} \right) \right),
\]

where $ h_v^l $ i the embeddings of nodes $ v $ and $ u $ at the $ l $-th layer, respectively. The aggregation function ($\text{AGG}(\cdot)$) collects information from neighboring nodes, and the combination function ($\text{COM}(\cdot)$) integrates this information with the current node's embedding. The initial node embedding $ h_v^0 $ is initialized with the corresponding node attributes $ X_v $.  
For a complete representation of the entire graph $ G $, a READOUT function is used, conveyed as:
\[
h_G^l = \text{READOUT} \left( \{ h_v^l \mid \forall v \in \mathcal{V} \} \right),
\]
where the READOUT operation can be any permutation-invariant function, such as summation, mean, or maximum pooling.

\subsection{Databases}
A database is a structured repository designed for efficient information storage, manipulation, and retrieval. It solves the challenge of managing large amounts of structured and unstructured data by providing systems that ensure consistency, security, and scalability while supporting data storage, querying, and updating. Modern databases enable complex queries, transactional processing, and data analysis through well-defined schemas, indexing mechanisms, and optimization techniques. Over the past few decades, database research has significantly matured, leading to well-established theoretical foundations. This progress has fueled the development of robust database management systems~\cite{postgre,traditional-join-gauss,neo4j}.

\begin{figure*}[t!]
	\centering 
	\resizebox{\linewidth}{!}
	{
		\includegraphics[width=\linewidth]{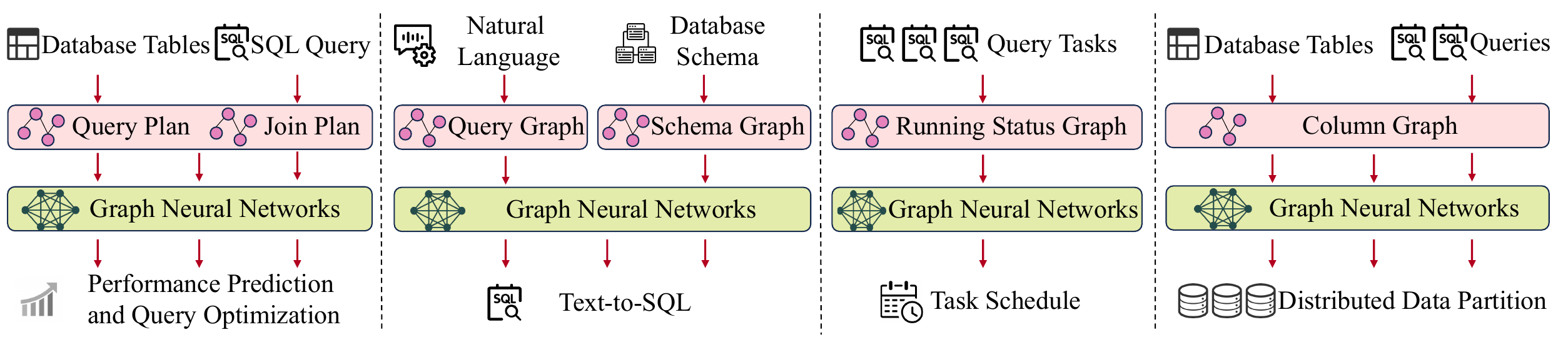}
	}
	\vspace{-0.2in}
	\caption{GNNs for Relational Databases. GNNs for relational databases typically represent element relationships or physical operations in relational databases as graphs, leveraging GNNs to generate corresponding representations for downstream tasks. GNNs may also serve as the models for these downstream tasks.}
	\vspace{-0.1in}
	\label{fig:traditional_db}
\end{figure*}

\subsection{Challenges}  
The application of GNNs to DB systems presents unique challenges due to graph-based message-passing mechanisms:  
\begin{itemize}  
    \item \textbf{Graph Representation of Database Problems.} 
    To harness the full potential of GNNs for addressing database-related challenges, it is crucial to efficiently encode the structured elements, constraints, and problems of databases into graph representations.

    \item \textbf{Feature Engineering for Database Contexts.}  
    Designing node and edge features tailored to database contexts is essential but demanding. Nodes should represent database entities, whereas edges should encapsulate relational semantics. Selecting relevant features for the constructed graph is crucial to address different problems.

    \item \textbf{GNNs' Capabilities in Databases.} 
    In database contexts, GNNs' capabilities must align with specific applications, such as query optimization or schema matching. The challenge lies in selecting or designing GNN architectures that balance computational efficiency with effectiveness. 
\end{itemize}  

In the following sections, we explore how GNNs contribute to each area and highlighting the differences among the various methods by discussing above challenges. Fig.~\ref{fig:taxonomy} presents a taxonomy that classifies existing GNNs for DB works into relational databases and graph databases.

\section{GNNs for Relational Databases}
At first, we explore how GNN-based approaches have been used to address different problems in relational databases. As shown in Fig.~\ref{fig:traditional_db}, GNNs for Relational DB encode element relationships or operations as graphs, generating representations or serving directly as models for downstream tasks.

\vspace{-0.07in}
\subsection{Performance Prediction}
Since GNNs excel at capturing relationships between related features (e.g., buffer size, computation costs), many works have leveraged GNNs to improve performance prediction.

Neo~\cite{traditionaldb-plan-Neo} and Bao~\cite{bao} convert logical and physical query plans into graph representations. They use GNNs to predict execution times. Particularly, Bao utilizes a vector tree to represent a query plan. Each tree node features a vector embodying the operator's one-hot encoding, cardinality, cost, and cache information. The execution time of these vector trees is then predicted using a tree convolutional neural network.
GPredictor~\cite{traditionaldb-performance-GPrecdictor} extends performance prediction to handle concurrent queries by extracting workload features that impact performance. It identifies physical operators in query plans, analyzes relationships such as data sharing and lock conflicts, and gathers database configuration parameters and active tasks. The concurrent query behavior is modeled as a graph, with nodes representing operators and edges capturing their relationships. GPredictor employs a graph-based learning model to generate embeddings and a graph prediction network to predict performance.
To enhance the generalization of performance prediction, Tosure~\cite{tradition-predict-tosure} uses Tree-LSTM to encode query plans, extracting transferable features independent of specific database instances. 

Additionally, some works focus on performance prediction in cloud database engines. 
BRAD~\cite{tradition-prediction-brad} improves performance prediction by using SQL text and query features to estimate query runtimes. It creates a feature graph through SQL statement parsing, incorporating logical features. The model utilizes a GNN with a unique query representation framework centered on these logical query attributes and data statistics, including estimated join selectivity.
Stage~\cite{tradition-predict-stage} employs a three-tiered strategy to address performance prediction. In particular, Stage's global model is constructed using a GCN. The model accepts a query's physical execution plan as a graph and predicts execution times. Each node in the input graph represents a physical operator.

The common framework for query performance prediction using GNNs transforms query plans into graph representations, where nodes and edges encode the relationships in the execution plan. GNNs learn feature representations from these graph structures to estimate query execution times or costs. They capture dependencies and interactions between query plan components, such as tables, indexes, and resources, to predict the overall impact of the performance.

\vspace{-0.07in}
\subsection{Query Optimization}  
Query processing plays a key role in ensuring the efficiency of database systems. 
This section explores how GNN-based approaches enhance query optimization tasks.

\subsubsection{Join Order Selection(JOS)}  
A query can have multiple join plans that yield the same result, but each plan incurs a different cost.
GNNs, with their ability to capture relationships between tables and columns, can optimize join orders to enhance query performance.

RTOS~\cite{traditionaldb-join-RTOS} uses Tree-LSTM to construct a join tree for a query in a reinforcement learning framework. During construction, Tree-LSTM combines information from columns and tables to estimate query execution costs as the long-term reward for a partial join tree. The join tree is incrementally expanded by adding one node at a time until all required join operators are incorporated. This approach effectively captures the join tree structure and adapts dynamically to schema changes.
JOGGER~\cite{traditionaldb-join-JOGGER} addresses the JOS problem by considering table connections. It first constructs a schema graph based on primary-foreign key relationships in the database and applies the DeepWalk algorithm to generate table representations. For each query, JOGGER creates an undirected query graph $ \mathcal{T}$. $\mathcal{T}$ is then encoded using GCNs along with the table representations to capture the relationships among the participating tables and their connections.

Contrary to the assumption in previous studies that each segment of a join tree equally impacts cost prediction, SOAR~\cite{traditionaldb-join-soar} accounts for the varied influence of different components. It resolves the JOS issue using a GAT. Firstly, it utilizes GAT to encapsulate the join trees' structure. Secondly, SOAR leverages GAT to highlight the differing impacts of join tree components on long-term rewards.

LOGER~\cite{traditionaldb-plan-LOGER} leverages the knowledge of the DBMS optimizer to refine the JOS search space. A query is naturally represented as a join graph, where nodes correspond to tables and edges denote join predicates. To address the JOS problem, LOGER employs a Graph Transformer to facilitate information exchange between adjacent table nodes while embedding structural information into each node. This approach effectively captures relationships between tables and join predicates, improving optimization quality.

Specific GNN models optimized for certain databases can enhance JOS performance.
RGOS~\cite{traditional-query-RGOS} introduces a join order selection algorithm specifically designed for TiDB. RGOS integrates Tree-LSTM and GCN with TiDB's architecture to represent states in the reinforcement learning environment. The Tree-LSTM encodes the existing subplan tree, while the GCN encodes the new subplan when it joins with the remaining tables, capturing graph relationships. These graph convolutional features are then input into the reinforcement learning Q-network to guide decision-making.

Existing frameworks for addressing the JOS problem with GNNs involve representing join orders as graph structures, where nodes correspond to tables or columns, and edges represent join predicates or relationships. GNNs are used to learn representations of these join trees or graphs, capturing complex interactions and dependencies. This process is often integrated with reinforcement learning techniques.

\subsubsection{Cardinality Estimation} 
Cardinality estimation (CE) is a critical component of database query optimization, tasked with predicting the number of rows (or tuples) a query or its specific components will return.

To improve CE precision, AutoCE~\cite{traditionaldb-performance-AutoCE} introduces a model advisor that quickly selects a suitable CE model based on the data set and specified metrics. 
In AutoCE, dataset characteristics are represented as a feature graph, such as column correlation, skewness, and domain size. Nodes in the graph encapsulate the features of individual tables, while the edges represent joins between tables. During model selection training, AutoCE employs a GIN to encode these feature graphs, facilitating accurate and efficient model recommendations.
For distributed relational databases, GCE~\cite{tradition-predict-GCE} introduces a novel architecture tailored to TiDB. GCE employs GNNs to create query plan representations specifically for CE tasks.
In this approach, Tree-LSTM is utilized to capture the structural features of the query plan tree, while a GCN is used to extract the relationships and connection topology in the query plan.

\subsubsection{Materialized View} 
Materialized Views (MVs) operate akin to standard tables. Precomputed and stored query results offered by MVs notably enhance query performance.
GnnMV~\cite{traditionaldb-View-GnnMV} tackles dynamic materialized view management by representing dynamic query workloads as a query graph. A GNN model is trained on this graph, taking node features as input and outputting benefit estimations for using an MV to answer a query. Key query features (e.g., operator types, metadata, and predicates) are encoded into the GNN.

\subsection{Text-to-SQL} 
SQL, as the query language in databases, continues to be a critical area. One notable focus is Text-to-SQL, which enables users to query databases using natural language, improving accessibility and usability~\cite{DBLP:journals/corr/abs-2408-05109,DBLP:conf/sigmod/Luo00CLQ21}. 
GNNs model structural relationships in databases, aligning database elements with natural language.

In GNNSQL~\cite{sql-bogin}, the database schema is represented as a graph using GNNs, including tables, columns, and key relationships. This novel method employs a RGCN to compute global node representations, which are then integrated into an encoder-decoder parser. This enables the conversion of natural language questions into SQL queries. 
Building upon GNNSQL, GLOBAL-GNN~\cite{sql-globalgnn} integrates global schema reasoning and database constants. It leverages GNNs in three distinct ways: (1) for relevance scoring, it uses a gated GCN to identify database constants likely to feature in the query; (2) for representation learning, a GCN calculates representations for database constants, used by the decoder to generate candidate queries; and (3) for candidate re-ranking, a re-ranking GCN selects the most appropriate query from the generated candidates.
IST-SQL~\cite{sql-ist-sql} addresses the Text-to-SQL problem in multi-turn scenarios.
ISR-SQL constructs a schema-state graph that captures relationships between tables and previously generated SQL keywords associated with the schema. 
Then, it utilizes an RGNN to encode schema-state representations for further SQL generation.

Previous work on text-to-SQL problems often overlooked the role of edge information in representation generation. 
LGESQL~\cite{sql-lgesql} transforms the original node-centric schema graph into an edge-centric graph to better model the edge topology.
LGESQL employs an RGAT to enhance the representations of both node-centric and edge-centric graphs.
SADGA~\cite{sql-sadga} points out the limitation of treating questions as sequences while representing database schemas as structured graphs. SADGA bridges this gap with a unified Gated Graph Neural Network (GGNN) encoder that cohesively models both the question and the schema. 
To further improve generalization capabilities, ShadowGNN~\cite{sql-shadowgnn} tackles the challenge of limited cross-domain generalization caused by schema semantics. It employs a Graph Projection Neural Network built on RGCN to abstract natural language queries and semantic schemas.
Similarly, TTFSQL~\cite{sql-tffsql} integrates a cross-graph attention mechanism into RGAT, allowing the model to dynamically adjust the attention weights between different graphs.

\begin{figure*}[t!]
	\centering 
	\resizebox{\linewidth}{!}
	{
		\includegraphics[width=\linewidth]{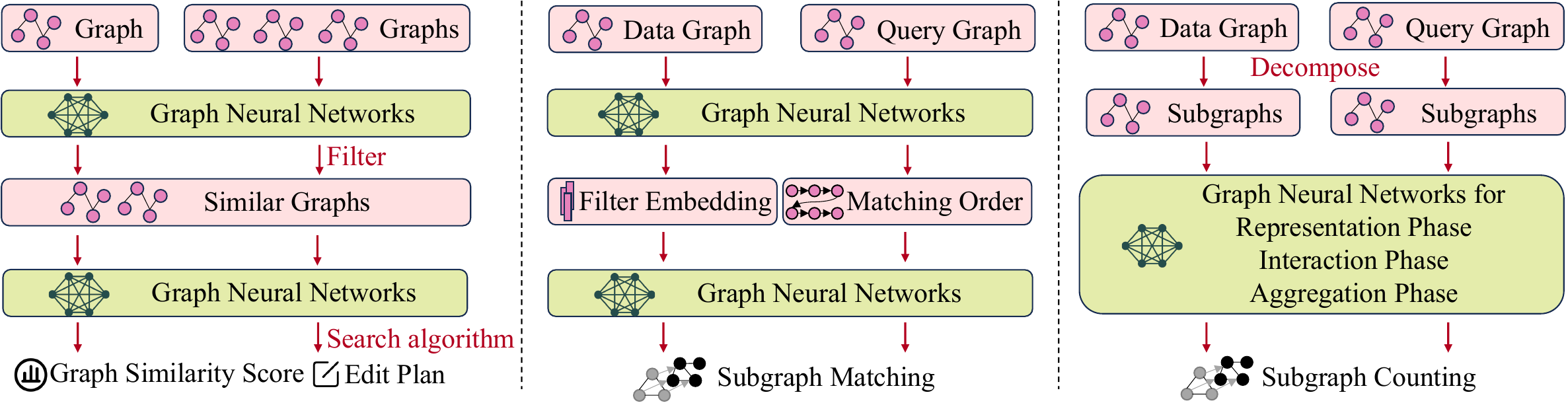}
	}
	\vspace{-0.2in}
	\caption{GNNs for Graph Databases. GNNs for graph databases typically transform graph structures into representation space for comparison or use these representations to accelerate traditional algorithms.}
	\label{fig:graphDB}
\end{figure*}

With the rise of pre-trained language models (PLMs) in various fields~\cite{DBLP:conf/edbt/QinL0018,DBLP:conf/sigmod/LuoQ00W18}, researchers have begun integrating PLMs with GNNs to address Text-to-SQL problems.  
ISESL-SQL~\cite{sql-ISESL-SQL} and  S$^2$SQL~\cite{sql-s2sql} uses PLMs to construct schema-linking graphs. They employ a modified RGAT to learn joint embeddings for questions and schema nodes.  Similarly, Graphix-T5~\cite{sql-t5} combines the T5 pre-trained language model with a semi-pretrained RGNN to integrate semantic information and establish schema linkage relationships.  
GRL-SQL~\cite{sql-GRL-SQL} also merges PLMs with GNNs, focusing on disconnected nodes in the question-schema graph.

\subsection{Others Problems in Relational Databases}
\textbf{Scheduling. } Optimal scheduling enhances the computational effectiveness of a database. 
Decima~\cite{tradition-predict-Decima} begins by encoding ongoing job stages and their dependencies as directed acyclic graphs (DAGs). These DAGs are processed using a GCN to generate the corresponding representations. The representations include information such as job stage attributes and the dependency structure of the DAG. \\
\textbf{Data Partition. } 
A well-designed partition improves database performance by allowing queries to access the relevant data efficiently.
Grep~\cite{traditional-partition-grep} improves the quality of data partition by introducing GNN-based technology. It begins by creating a column graph, where columns are vertices and joins are edges. Each vertex has a feature vector. Grep then uses attention mechanisms to select key columns for partitioning. A GNN is applied to capture global structures for each vertex.
A classifier is used to choose the best partition keys, ensuring optimal partitioning decisions. To avoid the cost of actual partitioning, Grep uses a graph learning model to predict database performance. \\
\textbf{Index Selection.}  
Indexes are essential for accelerating database queries.
SmartIndex~\cite{tradition-predict-smartindex} addresses this issue by using a GCN-powered cost estimation model for selecting workload-specific indexes. The process begins with an LSTM model that encodes nodes in query plans, where each node represents a specific DBMS operation. At the same time, SmartIndex traverses the query plan to get its query plan's adjacency matrix. The GCN processes the LSTM outputs and adjacency matrix to generate feature representations.
A ResNet model is then used to estimate the cost of each query plan and index pair.

\subsection{Summary of Methods on Relational Databases}
As shown in Fig.~\ref{fig:traditional_db}, the above methods significantly enhance the efficiency of the relational database by modeling database operations and relationships as graphs.
One approach involves representing database actions, such as query plans and join operations, as graphs. In these graphs, nodes correspond to operators or intermediate results, and edges represent dependencies or execution sequences. GNNs are used to encode these actions into detailed representations that can predict key metrics, including performance and execution cost. Such predictions enable query optimizers to select the most efficient execution plans.
Another approach focuses on modeling relationships in relational databases. Schema graphs capture the connections between columns and tables, reflecting data dependencies and structural relationships. Similarly, workload graphs describe the dependency structures of jobs, emphasizing shared resources and execution priorities. By processing these relationships, GNNs generate representations that support downstream tasks that contribute to more efficient database operations.

\vspace{-0.10in}
\section{GNNs for Graph Databases}
Graph databases store, query, and analyze data in graph form. This representation provides significant opportunities for leveraging GNNs power. As shown in Fig.~\ref{fig:graphDB}, this survey emphasizes graph database tasks, including graph similarity computation and graph query.

\vspace{-0.10in}
\subsection{Similarity Computation}
Graph similarity computation is a fundamental operation in graph databases. Since GNNs can effectively transform graphs into embeddings and extract their features, many studies have integrated GNNs into graph similarity computing problems.

\subsubsection{Graph Edit Distance (GED)}  
GED is widely used to identify a set of graphs from a graph database that are structurally similar to a given query graph and compute the edit distances between them.
GNNs excel at transforming graphs into embeddings, enabling efficient comparisons in the embedding space. Furthermore, GNNs can help traditional algorithms in the GED problem, such as A$^*$, by guiding them to better search directions.
In this section, we discuss these methods in detail. 

SimGNN~\cite{graphdb-gsc-gsc-simgnn} reformulates graph similarity computation as a learning problem. It starts by using GCNs to generate node embeddings for each graph, encoding both node features and structural properties. Subsequently, SimGNN calculates graph similarity through interaction between graph embeddings, yielding an graph similarity score.
At the same time, GMN~\cite{graphdb-GMN} introduces graph matching networks to improve verification accuracy. It employs a cross-graph attention mechanism to associate nodes across graphs, identifying structural differences to compute similarity scores.

Previous approaches often use fixed-length vectors to represent graphs, but real-world graphs typically vary in size. To address this, GraphSim~\cite{graphDB-gsc-ged-graphsim} directly performs neural operations directly on two sets of node embeddings. It employs a GCN to generate node representations and introduces a multiscale framework that utilizes outputs from multiple GCN layers to construct similarity matrices.
However, ERIC~\cite{graphDB-gsc-gsc-areg} challenges the necessity of computationally expensive node-to-node matching. It proposes a streamlined regularization technique, Alignment Regularization (AReg). During training, AReg enforces a node-graph correspondence constraint in the GNN encoder. At inference, only the graph-level representations are used to compute similarity scores. MGMN~\cite{graphDB-gsc-gsc-MGMN} combines a node–graph matching network for capturing interactions and a siamese GNN for learning global-level interactions.

Although many works improved efficiency and accuracy for GED, they required scanning the entire graph database. To address this, GHashing~\cite{graphdb-gsc-gsc-ghashing} introduces GNN-based semantic hashing for efficient GED computation on large datasets. It trains a GNN to generate graph embeddings, which are then converted into hash codes by an MLP to represent graph types, enabling constant-time lookups.
Similarly, GREED~\cite{graphDB-gsc-ged-greed} employs an indexing strategy to enhance the pruning process's efficiency. It uses a Siamese GIN to generate independent embeddings for graphs in a pair-independent manner. 

To better capture the relationships in graphs, 
H$^2$MN~\cite{graphdb-gsc-gsc-h2mn} introduces a novel approach by transforming each graph into a hypergraph to capture non-pairwise relationships.
Then, H$^2$ MN uses a hypergraph neural network to model complex structural relationships and achieve superior performance.
For explainability,
GOTSim~\cite{graphDB-gsc-gsc-graphotsim} generates node embeddings across multiple GCN layers and derives the optimal transformation cost. The final similarity score is aggregated from these costs across all layers.

Search-based works address limitations of prior learning-based methods cannot recover the actual edit path.
Noah~\cite{graphdb-gsc-ged-noah}, GENN-A$^*$~\cite{graphdb-gsc-ged-GENN-A*} and MATA$^*$~\cite{graphDB-gsc-ged-mata*} focus on guiding the search directions of the A$^*$ algorithm. Noah uses GIN to learn the cost estimation function $h(\cdot)$. GENN-A$^*$ introduces the Graph Edit Neural Network (GENN). GENN consists of GCN and SplineCNN to generate node embeddings. These embeddings are cached for further dynamic graph embedding during the A$^*$ search process. 
MATA$^*$ reframes GED computation as a node-matching problem. It uses a structure-enhanced GNN (SEGcn) to identify candidate matching nodes, effectively pruning unpromising search directions in the A$^*$ algorithm.
GEDGNN~\cite{graphdb-gsc-ged-gedgnn} employs a two-step process to recover the edit path. First, it predicts the GED value and a matching matrix using GNNs. Second, the k-best matching algorithm generates multiple node matching from this matrix to derive the edit path.


\vspace{-0.1in}
\subsection{Graph Query}
Subgraph matching and subgraph counting involve identifying all subgraphs in a data graph that match a given query graph or determining the count of such subgraphs. Traditional algorithms face significant scalability challenges due to the tasks' NP-complete nature.
GNN-based methods enable more scalable solutions for subgraph matching and counting.

\vspace{-0.1in}
\subsubsection{Subgraph Matching}
GNN-based methods for subgraph matching can be broadly divided into two categories: learning-based approaches that rely entirely on GNNs and hybrid approaches that combine GNN techniques with traditional heuristic search methods to assist in the matching process.

There are some learning-based subgraph matching works focus on approximate subgraph matching result. Neuromatch~\cite{graphDB-gq-sm-neuromatch} leverages a GNN to embed nodes in the data graph. For a given query graph, it selects a central node and maps the query graph into the embedding space using the same GNN. A data vertex is retained as a candidate only if its embedding is located in the top-right quadrant of the query graph's central node. This approach efficiently identifies candidate subgraphs likely to match the query graph. ISONET~\cite{graphDB-gq-sm-isonet} extends the scope of GNNs by proposing a framework incorporating context-sensitive representations for nodes and edges.

Hyper methods can be divided into two types. The first type uses GNN embeddings as an indexing mechanism. The second type leverages embeddings to generate better traditional matching orders.
For the first type, Duong~\cite{graphDB-gq-sm-chi} and GNN-PE~\cite{graphDB-gq-sm-gnnpe} use GNN-based embeddings to filter unpromising search branches. Duong introduces an indexing mechanism, mapping graphs into a numerical space where structurally similar nodes and subgraphs are close to each other. When a query graph is received, the search is narrowed to candidate regions near the query's embedding, reducing the search space.
GNN-PE employs GNN-based embeddings for graph paths. These embeddings define a dominant relationship, ensuring that paths with dominant embeddings maintain a subgraph relationship. This approach enables GNN-PE to prune candidates effectively and transform subgraph matching into a dominating region search in the embedding space.
For the second type, RL-QVO~\cite{graphDB-gq-sm-rlqvo} use GNNs to generate high-quality matching order. The process begins by generating feature vector for each node in the query graph, based on statistical heuristics such as node degree and label frequency. RL-QVO then employs GNNs to produce vector representations that capture comprehensive graph-level information.
These representations are used in a reinforcement learning framework to generate optimal matching orders.

Overall, GNNs for subgraph matching contribute to two key areas. First, they utilize GNNs to map graph relationships into a unified embedding space. These embeddings can directly support learning-based subgraph matching or act as auxiliary filters in hybrid methods. Second, GNNs capture graph features to generate higher-quality matching order.

\vspace{-0.05in}
\subsubsection{Subgraph Counting}
GNNs are employed in two primary directions for subgraph counting problem. 
First, they are used to generate better embeddings for data graphs and query graphs. Second, they are leveraged to enhance scalability, enabling subgraph counting for larger and more complex graphs.

NSIC~\cite{graphDB-gq-sc-rgin} treats the data graph as an information source and the query graph as a retrieval query. It employs RGCN/RGIN to learn representations from adjacency matrices and vertex features. These representations are processed through an interaction module to capture correlations and are further combined with size information to predict subgraph counts.
At the same time, \cite{graphDB-gq-sc-mpnns} demonstrates that 2-invariant Graph Networks can effectively count subgraphs for star-shaped patterns.
To capture fine-grained structural information, Count-GNN~\cite{graphDB-gq-sc-countgnn} introduces a subgraph counting framework based on an edge-centric GNN that propagates and aggregates messages specifically for edges. 

Several studies aim to improve the efficiency of subgraph counting in large graphs by decomposing the data graph, the query graph, or both.
LSS~\cite{graphDB-gq-sc-alss_sig} decomposes query graphs and introduces a subgraph counting framework based on sketch learning. This method first extracts features for each substructure from the query graph as a fixed-length vectorized representation. Specifically, ALSS employs GNNs to represent substructures. It then uses a self-attention mechanism to learn an aggregation function, followed by an MLP to predict the final count. 
On the contrary, NeurSC~\cite{graphDB-gq-sc-neursc} extracts substructures from the data graph. NeurSC uses an estimator for subgraph counting. The estimator comprises an intra-graph neural network, an inter-graph neural network, and a Wasserstein discriminator. Similarly, DeSCo~\cite{graphDB-gq-sc-DeSCo} divides the data graph into small neighborhoods, encoding their local information using a GNN with subgraph-based heterogeneous message passing.
LearnSC~\cite{graphDB-gq-sc-learnsc} simultaneously decomposes both the query graph and data graph. Additionally, LearnSC presents a comprehensive subgraph counting framework with five phases: decomposition, representation, interaction, estimation, and aggregation. LearnSC uses GIN to acquire each node's representation in the representation phase and employs GIN as the base network in a cross-graph learning model during the interaction phase. 

\vspace{-0.05in}
\subsection{Other Problems in Graph Databases}

\textbf{Graph Partitioning.} 
Graph partitioning plays a critical role in graph database. GCNSplit~\cite{graphDB-others-partition-gcnsplit} introducing a novel framework designed for graph partitioning in streaming environments. It utilizes an immutable, fixed-size model to capture the characteristics of the graph stream.
GCNSplit leverages GCN to generate vertex embeddings and enhances the model with load constraints and assignment heuristics. \\
\textbf{Knowledge Graph.}
Knowledge graphs are pivotal to graph databases. This paper concentrates on database-specific facets such as cardinality estimation.
Knowledge graphs can be queried using SPARQL, where conjunctive queries require graph pattern matching. GNCE~\cite{kg-ce-gnce} leverages GNNs to model the query structures of knowledge graphs. It generates RDF2Vec embeddings, trains a GNN to minimize cardinality discrepancies, and predicts cardinalities for new queries.
HRQE~\cite{kg-ce-hrpe} addresses hyper-relational knowledge graphs, using a GNN-based model that incorporates qualifier information. It employs a pre-trained mechanism for inferring qualifier features and enhances the GNN architecture with additional layers. HRQE refines the representations through a linear projection vector and an MLP decoder to predict query cardinality.

\vspace{-0.05in}
\subsection{Summary of Methods on Graph Databases}
GNN-based methods often transform traditional graphs into representations, effectively capturing essential features. These representations can be broadly categorized into three main types.
The first category involves comparing graphs in the representation space. This capability is particularly useful for similarity-based queries, where graphs are matched based on proximity in the embedding space. Additionally, these representations play a crucial role in filtering operations. For instance, in subgraph matching, the representations are used to prune unpromising candidate graphs early in the process.
The second category leverages these representations as an auxiliary tool to enhance traditional graph algorithms. For example, GNN-generated representations guide A$^*$ algorithms by providing heuristic search directions. Similarly, in subgraph matching problems, these representations are instrumental in generating high-quality search plans. 
The third category focuses on utilizing these representations to enhance the functionality and performance of graph databases. For example, GNN representations improve data partitioning by capturing structural nuances. Moreover, by accurately modeling the relationships between entities and their attributes, GNN-based methods provide precise cardinality estimation.

\vspace{-0.07in}
\section{Future Directions}
The study of GNNs for databases is advancing quickly, offering numerous challenges and opportunities. Here, we explore some of them.
\\
\textbf{Database Implementation.} 
While many GNN-based methods aim to improve database efficiency and capability, few have been directly implemented in business databases.  Future work could explore optimizing GNN-based techniques tailored to specific database frameworks. 
Moreover, implementing GNN-based methods in cloud databases represents a significant opportunity for future research.
\\
\textbf{Relational Databases Problems based on Graph.} 
As shown in Fig.~\ref{fig:traditional_db}, many GNN-based methods for relational databases rely on transforming database schemas and their underlying relationships into graph representations. By representing database elements, like tables and columns, and relationships, like graph nodes and edges, researchers can unlock new possibilities for addressing a wide range of database problems.
\\
\textbf{Integration with Large Language Models.} 
The combination of large language models (LLMs) and graph learning methods has emerged as a significant area of research.~\cite{DBLP:conf/cidr/0001YF0LH24,DBLP:journals/pvldb/XieLLT24}.
LLMs excel at understanding and processing unstructured textual data, while GNNs provide powerful tools for modeling structured relationships and dependencies in graph-structured data. By combining the strengths of both paradigms, researchers can address a broader spectrum of database challenges, from natural language query understanding to optimizing graph-based queries and enhancing data retrieval accuracy.

\vspace{-0.06in}
\section{Conclusion}  
Graph Neural Networks (GNNs) have demonstrated remarkable potential in the database domain by capturing features and connections across various components. Numerous studies have contributed to this interdisciplinary area, aiming to enhance the capabilities and efficiency of both relational and graph databases. In this survey, we first introduce some key concept in this joint area. Then, we review these efforts in this survey by categorizing them into relational and graph database contexts. For each category, we highlighted the specific problems addressed and the roles GNNs played in solving them. Additionally, we outlined several promising directions for future research in the intersection of databases and GNNs.
\clearpage
\footnotesize
\bibliographystyle{named}

\end{document}